\documentclass[reprint,superscriptaddress,aps,prd,floatfix,nofootinbib,bibnotes]{revtex4-1}
\usepackage[colorlinks=true,linkcolor=black,citecolor=blue, urlcolor=blue,bookmarks=false]{hyperref}
\hypersetup{breaklinks=true}
\usepackage[utf8]{inputenc}
\usepackage{natbib}
\usepackage{bm,mathtools,cleveref}
\usepackage{array}
\usepackage{epsfig}
\usepackage{amsmath}
\usepackage{amssymb}
\usepackage{multirow}
\usepackage{graphicx,subcaption}
\usepackage{comment}
\usepackage{enumitem}
\graphicspath{{./fig/}}
\usepackage[normalem]{ulem}  
\usepackage[x11names]{xcolor} 

\renewcommand{\sout}{\bgroup \color{red} \ULdepth=-.5ex \ULset}

\begin{document}
\title{Long-Range 
$N-J/\psi$ 
Interaction from an Operator Product Expansion Perspective
}

\author{Seokwoo Yeo}
\email{ysw351117@gmail.com}
\affiliation{Department of Physics and Institute of Physics and Applied Physics, Yonsei University, Seoul 03722, Korea}
\author{In Woo Park}
\email{darkzero37@naver.com}
\affiliation{Department of Physics and Institute of Physics and Applied Physics, Yonsei University, Seoul 03722, Korea}

\author{Su Houng Lee}%
\email{suhoung@yonsei.ac.kr}
\affiliation{Department of Physics and Institute of Physics and Applied Physics, Yonsei University, Seoul 03722, Korea}

\date{\today}
\begin{abstract}

A recent lattice QCD study has shown that the $N-J/\psi$ potential is attractive at all distances, and its long-range tail is well described by two-pion exchange. 
Here, we study to what extent the long-range part of the attraction can be reproduced from the perspective of the operator product expansion (OPE). 
This is accomplished by extracting the leading-order four-quark operator that couples to two pions and calculating its contribution to the $J/\psi$ mass in nuclear matter, to linear order in density, within the QCD sum rule framework.
Using previous estimates of the four-quark operators for the chiral symmetric and breaking parts, we obtain a mass decrease that is smaller in magnitude but qualitatively consistent with the attraction obtained in the lattice QCD calculation. By expressing the interaction in terms of four-quark operators, we can analyze the effects of chiral symmetry restoration in dense matter on the masses of the $J/\psi$ and other mesons composed of heavy quarks.

\end{abstract}

\maketitle

\section{Introduction}

Understanding how the properties of the vector mesons are modified in nuclear matter has long been a central objective in nuclear and hadronic physics \cite{PISARSKI1982,Hatsuda1992,Rapp2009,OHNISHI2020}.
Such studies provide access to nonperturbative quantum chromodynamics (QCD) at finite density, shedding light on phenomena such as partial restoration of chiral symmetry \cite{Ejima2025} and alterations in long‑range gluonic correlations.

Charmonium-nucleus interaction has also been of great interest for many years \cite{Brodsky:1989jd,Klingl:1998sr,Lee:2000csl}.   
The $J/\psi$ is a clean probe: composed of a charm–anticharm pair, it interacts with a nucleon mainly through multiple gluon exchange \cite{Brodsky:1989jd,Klingl:1998sr,Morita2008}.
Consequently, any in-medium mass shift or width broadening of $J/\psi$ reflects modifications of the gluon configurations in nuclear matter, in contrast to how light vector mesons are sensitive probes of chiral symmetry restoration in nuclei \cite{Hatsuda1992,Koike1997,Rapp2009,Lee:2024yyf}.

Recent lattice QCD calculations have shown that the nucleon–$J/\psi$ interaction is attractive at all distances, with the long distance part dominated by two-pion exchange (TPE) \cite{Lyu2022,Lyu2025}. This result shows that, while the attraction arising from gluon exchange is important at short distances, the overall attraction in matter receives a nontrivial contribution from two-pion exchange.
The resulting S‑wave scattering length is about 0.3–-0.4 fm \cite{Lyu2025}, roughly an order of magnitude larger than values inferred from earlier estimates based on $J/\psi$ photoproduction within vector meson dominance models \cite{Gryniuk2016,Strakovsky2020,Pentchev2021}. The lattice result is consistent with the dispersive analysis of Wu et al.\cite{Wu2025}, which finds an attractive scattering length with magnitude $|a_{NJ/\psi}|\gtrsim0.16~\text{fm}$ from the soft-gluon exchange; its long-range part is dominated by two-pion exchange. The non-trivial contribution from the two-pion exchange implies that the chiral symmetry breaking effects can influence heavy quark systems or mesons composed of quarks other than $u$ and $d$.

To study the properties of the $J/\psi$ in nuclear matter, QCD sum rules (QSR) have previously been employed \cite{Klingl1999,S.S.Kim2001}. Since the matrix elements in the QCD sum rule approach in nuclear matter are calculated to leading order in density, the resulting mass shift can be related to the scattering with a nucleon and thus to the scattering length \cite{Hatsuda:1995dy,Hayashigaki1999}.
In earlier works, the QCD condensates included were the scalar gluon condensate and the twist-2 gluon matrix elements at dimension-4. Later studies incorporated contributions from scalar and tensor operators at dimension-6. All these QSR analyses consistently predict a modest downward mass shift of about 5–10 MeV at normal nuclear density \cite{Klingl1999,Hayashigaki1999,S.S.Kim2001}.

In this work, we use QSR to investigate the in-medium mass shift of the $J/\psi$ meson at normal nuclear matter density arising from the long-range two-pion exchange contribution.
To isolate the pion-coupled component of the four-quark condensate, we first identify the leading order contributions from the four-quark operator. Then, we apply the Fierz transformation to extract the two-pion contribution. Finally, making use of previous results on the matrix elements of the four-quark operators \cite{Kim:2020zae,Kim:2021xyp,Lee:2023ofg}, we estimate the density‑dependent part of the pion‑associated operators and study their effect on the $J/\psi$ mass in nuclear medium. As we identify the important four-quark operators, we can study its structure to clarify the origin of the medium effects coming from chiral properties of the medium.

\section{QCD Sum Rules}
\label{qsr}
To calculate the mass shift of the $J/\psi$, we start with the two current correlator,
\begin{align}
    \Pi_{\mu\nu}(q)=i\int d^{4}x e^{iqx}\langle T\{j_{\mu}(x)j_{\nu}(0)\}\rangle,
\end{align}
where the current is $j_{\mu}=\bar{c}\gamma_{\mu}c$.
We then define the dimensionless correlator \cite{Morita2010}
\begin{align}
    \tilde{\Pi}(q^2)=-\frac{\Pi_{\mu}^{\mu}(q)}{3q^2}.
\end{align}
We take $\vec{q}=0$ and study the mass shift at rest so that there is no difference between the longitudinal and transverse component \cite{Lee:1997zta}. 
Using the analytic structure of the dispersion relation, the real and imaginary parts of the correlator are related by
\begin{align}
    \text{Re}\Pi(q)=\frac{1}{\pi}\int^{\infty}_{0}ds\frac{\text{Im}\Pi(s)}{s+Q^2}, \label{dispersion}
\end{align}
with $Q^2=-q^2$. To suppress contributions from excited resonances, continuum, and high dimension condensates, we apply the Borel transform, defined as
\begin{align}
    \mathcal{M}^{V}_{OPE}(M^{2})=\lim_{\substack{n,Q^{2} \to \infty,\\{Q^{2}/n=M^{2}}}}(Q^{2})^{n+1}\frac{1}{n!}\left (-\frac{d}{dQ^{2}}\right )^{n}\Pi(Q^2),
\end{align}
where $M^2$ is the Borel mass.

We evaluate the correlator using the operator product expansion (OPE), retaining operators up to mass dimension 6 and working at leading order in $\alpha_s$. 
The contribution of scalar gluon operators up to dimension 6 was reported in \cite{Reinders1981,bertlmann1982,nikolaev1983}, where two independent gluon operators were identified. The Wilson coefficients of operators with spin indices, which contribute in the medium or in the presence of external hadrons, have been calculated up to dimension-6 \cite{S.S.Kim2001}. Here, there are three independent operators. 
In this study, we estimate the dominant contribution that couples to two-pion states up to dimension-6 operators.  Accordingly, we retain the vacuum contributions to dimension-6, 
which are the dimension-4 gluon condensate and the two independent scalar operators at dimension-6.
For the density-dependent part, we include only the operators that couple to two-pion states in order to estimate the in-medium mass shift induced by two-pion exchange.
The relevant operators are listed in Table \ref{tab:operators}.

Now, consider calculating the quark operators that couple to two-pion states. Since the lowest Fock component of a pion consists of a quark and an antiquark, the leading operators that couple to the two-pion states are four-quark operators composed of light quarks. The light quark pair comes through the gluon field. If the gluon field carries soft momentum, one can convert it to the quark-antiquark operator.
This is accomplished by applying the equation of motion to the gluon operator $D_{\mu}G^{a}_{\alpha\mu}=g\sum_{i}\bar{q}_{i}\gamma_{\alpha}\frac{\lambda^{a}}{2}q_i$. Therefore, one obtains a four-quark operator from the relevant dimension-6 gluonic operator, as shown in Table \ref{tab:operators}.
Four-quark operators that are obtained this way will contribute to the leading term. On the other hand, if the gluon field carries a hard momentum, it will couple to the quark-antiquark pair with one of the created quarks carrying the hard momentum, leaving only one quark field to be soft. Therefore, one needs another hard light quark, which will also come from a hard gluon line. Therefore, quark operators coming from hard gluons inevitably involve higher order coupling and lead to four-quark operators of $\mathcal{O}(\alpha_{s}^{3})$. We keep only the leading $\alpha_s^{2}$ quark operator coming from soft-gluons.
The gluon operators that convert into four-quark operators and used in this work are listed in Table \ref{tab:operators}. 
Here, ‘$ST$’ denotes the operation that projects the Lorentz indices onto their symmetric traceless part.

\begin{widetext}
\begin{table*}[ht]
  \centering
  \renewcommand{\arraystretch}{1.5}
  
  \begin{tabular*}{0.7\textwidth}{@{\extracolsep{\fill}} l c c}
    \hline\hline
    Gluon Operators & \text{Dimension} & \text{Spin} \\ 
    \hline
    $\langle\frac{\alpha_s}{\pi}G^{a}_{\mu\nu}G^{a}_{\mu\nu}\rangle$ & 4 & 0 \\ 
    \hline
    $\langle \frac{\alpha_s}{\pi}gf^{abc}G^{a}_{\mu\nu}G^{b}_{\nu\lambda}G^{c}_{\mu\lambda}\rangle$
     &
    \multirow{2}{*}{6} & \multirow{2}{*}{0} \\ 
    $\langle \frac{\alpha_s}{\pi}G^{a}_{\alpha\mu}D_{\mu}D_{\nu}G^{a}_{\alpha\nu} \rangle$=$\langle -\alpha_{s}^{2}( \bar{q}_i\gamma_{\mu}\lambda^{a}q_i)^2\rangle$  & & \\ 
    \hline
    $\langle \frac{\alpha_s}{\pi}G^{a}_{\alpha\mu}D_{\mu}D_{\nu}G^{a}_{\beta\nu}|_{ST} \rangle$ = $\langle -\alpha_{s}^{2}\bar{q}_{i}\gamma_{\mu}\lambda^{a}q_{i}\bar{q}_{j}\gamma_{\nu}\lambda^{a}q_{j}|_{ST}\rangle$ &
    6 & 2 \\ 
    \hline\hline
  \end{tabular*}
  \caption{Operators used in this OPE analysis.}
  \label{tab:operators}
\end{table*}
\end{widetext}

Expanding the current correlator via the OPE and then performing the Borel transform yields \cite{bertlmann1982,nikolaev1983,S.S.Kim2001,Morita2010}:
\begin{align}
    \mathcal{M}&({\nu})=e^{-\nu} A^{V}({\nu})\left[1+a^{V}(\nu)\alpha_{s}(\nu)+b^{V}(\nu)\phi_{b}\right.\nonumber\\
    &\left.+s^{V}(\nu)\phi_{c}+ \left(s^{V}(\nu)+\frac{2}{3}t^{V}(\nu)\right)\phi_d+\tilde{y}^{V}(\nu)\phi_e \right], \label{ope}
\end{align}
where $\nu=4m^{2}_{c}/M^{2}$ and $m_{c}=1.262\,\text{GeV}$ \cite{Morita2010}.
The $\phi$'s are
\begin{align}
    &\phi_b\equiv \frac{4\pi^2}{9(4m^{2}_{c})^2}G_{0},\\
    &\phi_c\equiv \frac{1}{3\cdot 540(4m_{c}^{2})^3}G_{3},\\
    &\phi_d\equiv \frac{4\pi^2}{3\cdot 1080 (4m_{c}^2)^3}\langle -\alpha_{s}^{2}( \bar{q}_i\gamma_{\mu}\lambda^{a}q_i)^2\rangle, \label{scalar four quark}\\
    &\phi_{e}\equiv Y_{00}\;\;\text{with}\;\;Y_{\mu\nu}=\langle -\alpha_{s}^{2}\bar{q}_{i}\gamma_{\mu}\lambda^{a}q_{i}\bar{q}_{j}\gamma_{\nu}\lambda^{a}q_{j}|_{ST}\rangle.
\end{align}
All parameters are given at the renormalization scale of $\mu=1$ GeV, unless otherwise stated. The gluon condensates are given by $G_{0}\equiv \langle \frac{\alpha_{s}}{\pi}G^{a}_{\mu\nu}G^{a}_{\mu\nu}\rangle_0=0.012\,\text{GeV}^{4}$ \cite{Shifman1978,Shifman1979}, $G_{3}\equiv \langle g^{3}f_{abc}G^{a}_{\mu\nu}G^{b}_{\nu\lambda}G^{c}_{\lambda\mu}\rangle_0 =0.0467\,\text{GeV}^{6}$ \cite{nikolaev1983}. The quark condensate is $\langle \bar{u}u \rangle_{0}=(-0.246\,\text{GeV})^3$ \cite{Aoki2020}. The $\pi N$ sigma term $\sigma_{\pi N}\equiv 2m_{q}\langle N|\bar{q}q|N\rangle=39.7\,\text{MeV}$ \cite{Aoki2020}. The normal nuclear matter density is $\rho_0=0.17\,\mathrm{fm}^{-3}$. In the perturbative part, we obtain the strong coupling $\alpha_{s}(\nu)$ at scale $\nu$ by solving the running coupling equation from $\alpha_{s}(8m^{2}_{c})$=0.21 \cite{Morita2010}. The coupling constant in the vacuum part of the nonperturbative term is given by $\alpha_s(\mu=1~\text{GeV})=0.472$ \cite{Tanabashi2018,Kim2020}. In the density-dependent part of the four-quark condensate, because we are dealing with the long-range regime dominated by two-pion exchange, we adopt the infrared value of the strong coupling, $\alpha_{s,\text{IR}}$=0.7 \cite{nikolaev1983,S.S.Kim2001,Deur2016}. The Wilson coefficients $A^{V}(\nu)$, $a^{V}(\nu)$, $b^{V}(\nu)$, $s^{V}(\nu)$, and $t^{V}(\nu)$ are listed in Refs. \cite{nikolaev1983,S.S.Kim2001,Morita2010}. The newly computed term $\tilde{y}^{V}(\nu)$ is presented in Eq.(\ref{ytilde}).

For the vacuum sum rule for $J/\psi$, we use the vacuum saturation hypothesis for the entire four-quark condensate. After obtaining the vacuum mass, we evaluate the in-medium mass shift due to long-range two-pion exchange arising solely from the density-dependent four-quark operators, which couple to two-pion states.

Let us now consider those operators in detail.
To project out the pion-coupled component of the four-quark condensates from $\phi_d$ and $\phi_e$, we first perform a Fierz transformation and keep only the pseudoscalar and axial vector channels. The Fierz transformation of the spin-0 four-quark condensate $\langle \bar{q}_{i}\gamma_{\mu}\lambda^{a}q_{i}\bar{q}_{j}\gamma_{\mu}\lambda^{a}q_{j}\rangle$ is given by:
\begin{align}
    2\langle\bar{q}_i\gamma_5q_j\bar{q}_j\gamma_5q_i\rangle+\langle\bar{q}_i\gamma_5\gamma_{\mu}q_j\bar{q}_j\gamma_5\gamma_{\mu}q_i\rangle. \label{four-quark fierz}
\end{align}

The four-quark operator can be divided into chiral symmetric and chiral breaking parts \cite{Kim:2020zae,Kim:2021xyp}. Furthermore, using the OPE for the $\rho$ and $a_1$ mesons, which are chiral partners, and employing their vacuum properties, one can estimate the vacuum expectation values of both the symmetric and breaking parts of the four-quark operators composed of axial and vector currents.  
It was found that the matrix element of the chiral symmetry breaking operator is similar to that obtained using the vacuum saturation approximation. However, the matrix element of the chiral symmetric operator has a magnitude comparable to that of the breaking term—contrary to the vacuum saturation hypothesis \cite{Kim:2020zae,Kim:2021xyp}, which predicts it to vanish.  This leads to the so-called $\kappa$ parameter, which is introduced to correct the results obtained from the vacuum saturation hypothesis, and is typically taken to be larger than 1 in order to reproduce the vacuum meson masses. In nuclear matter,  intermediate states couple to nucleons, so the deviation from the exact vacuum saturation is expected to grow larger. To quantify this deviation, we introduce the parameter $\kappa$, whose values will be discussed in Sec.~\ref{sec:qsr_result}. Applying the vacuum saturation approximation with the parameter $\kappa$, the pion-coupled pseudoscalar and axial-vector four-quark matrix elements reduce to the following expressions:
\begin{align}
     \langle\bar{q}_i\gamma_5q_j\bar{q}_j\gamma_5q_i\rangle_{\mathrm{v.s.}}&=-\kappa\frac{1}{3}\langle \bar{u}u\rangle^2,\quad(q_i = u,d),\\
     \langle\bar{q}_i\gamma_5\gamma_{\mu}q_j\bar{q}_j\gamma_5\gamma_{\mu}q_i\rangle_{\mathrm{v.s.}}&=+\kappa\frac{4}{3}\langle \bar{u}u\rangle^2,
\end{align}
where the subscript v.s. indicates the vacuum saturated matrix element. Hence, the Fierz-projected spin-0 four-quark condensate becomes 
\begin{align}
    2\langle\bar{q}_i\gamma_5q_j\bar{q}_j\gamma_5q_i\rangle_{\mathrm{v.s.}}+\langle\bar{q}_i\gamma_5\gamma_{\mu}q_j\bar{q}_j\gamma_5\gamma_{\mu}q_i\rangle_{\mathrm{v.s.}}=\frac{2}{3}\kappa\langle \bar{u}u\rangle^2. \label{scalar}
\end{align}
Here, the density-dependent part of the condensate $\langle\bar{u}u\rangle^2$ is 
\begin{align}
    \langle\bar{u}u\rangle^{2}_{\Delta\rho}\equiv&\langle\bar{u}u\rangle^{2}_{\rho}-\langle\bar{u}u\rangle^{2}_{0}\nonumber\\
    \simeq&\langle\bar{u}u\rangle^{2}_{0}\left(\left(1-a\frac{\rho}{\rho_0}\right)^2-1 \right) \label{linear density}\\
    =&-1.38\cdot 10^{-4}\text{GeV}^6\nonumber
\end{align}
at $\rho=\rho_0$ with 
\begin{align}
    a=-\frac{\rho_0\sigma_{\pi N}}{2m_{q}\langle\bar{u}u\rangle^{2}_{0}}=0.385.
\end{align}
In Eq.(\ref{linear density}), we take the linear density approximation for the quark condensate.
\begin{align}
    \langle \mathcal{O}\rangle_{\rho} \simeq \langle \mathcal{O}\rangle_0+\rho\langle \mathcal{O}\rangle_{N}.
\end{align}

The spin-2 four-quark condensate is transformed as (see Eq.(\ref{nonscalar}))
\begin{align}
    &\langle\bar{q}_{i}\gamma_{\mu}\lambda^{a}q_{i}\bar{q}_{j}\gamma_{\nu}\lambda^{a}q_{j}|_{ST}\rangle\\
    &\rightarrow \frac{1}{4}g_{\mu\nu}\langle \bar{q}_i\gamma_5\gamma_{\mu}q_j\bar{q}_j\gamma_5\gamma_{\mu}q_i\rangle-\langle \bar{q}_i\gamma_5\gamma_{\mu}q_j\bar{q}_j\gamma_5\gamma_{\nu}q_i\rangle .\label{nonscalarfierz2}
\end{align}
Applying vacuum saturation to Eq.(\ref{nonscalarfierz2}) in nuclear matter yields only a contribution $\tfrac{9}{8}\kappa\rho_{0}^{2}$, which is numerically negligible at saturation density.

In the imaginary part of the dispersion relation given in Eq.\eqref{dispersion} we use the  pole+continuum ansatz for the spectral function and write
\begin{align}
    \rho^{V}(s)=f_0\delta(s-m^{2}_{J/\psi})+\frac{1}{\pi}\text{Im}\tilde{\Pi}^{V,\text{pert}}(s)\theta(s-s_0),
\end{align}
where $f_0$ is a pole residue and $s_0$ is a continuum threshold. For the explicit form of Im$\bar{\Pi}^{V,\text{pert}}(s)$, see Refs. \cite{Reinders1985,Morita2010}.

After performing the Borel transform on Eq.(\ref{dispersion}) and taking the ratio with its derivative, we obtain the following expression for the mass:
\begin{align}
    m_{J/\psi}(M^2,s_0)=\sqrt{-\frac{\frac{\partial}{\partial(1/M^2)}\bar{\mathcal{M}}^{V}(M^2,s_0)}{\bar{\mathcal{M}}^{V}(M^2,s_0)}},
\end{align}
where $\bar{\mathcal{M}}^{V}(M^2,s_0)=\mathcal{M}^{V}_{OPE}(M^2)-\int^{\infty}_{s_0}ds e^{-s/M^2}\rho^{V}(s)$.

In order to identify a reliable Borel mass region, we choose the following Borel window:
\begin{align}
    M_{min}^2:&\left| a^{V}(\nu)\alpha_{s}(\nu)+b^{V}(\nu)\phi_b+s^{V}(\nu)\phi_c\right.\nonumber\\
    &\left.\left(s^V(\nu)+\frac{2}{3}t^V(\nu)\right)\phi_d+\tilde{y}^V(\nu)\phi_e\right|\leq 0.3 \\
    M_{max}^2:&\left| \frac{\int^{\infty}_{s_0}dse^{-s_0/M^2}\rho^{V}(s)}{\mathcal{M}^{V}_{OPE}(M^2)}\right|\leq 0.4
\end{align}
$M_{min}^2$ is determined to guarantee the convergence of the OPE series and $M_{max}^2$ is set to maintain the pole dominance. We refer to the $\chi^2$ adjustment \cite{Kim2020} for the optimization procedure.

\section{Results and Discussion}
\label{result}
\subsection{Two-pion Exchange Potential}

We first estimate the magnitude of attraction coming from two-pion exchange using the lattice result.
Following the lattice QCD analysis \cite{Lyu2025}, which fitted the $N-J/\psi$ interaction with a long-range TPE potential, we adopt the potential,
\begin{align}
    V(r)=-\alpha\frac{e^{-2m_{\pi}r}}{r^2}, \label{tpe potential}
\end{align}
with potential strengths
$\alpha^{(S=3/2)}_{J/\psi}=22, ~\alpha^{(S=1/2)}_{J/\psi}=23~\text{MeV}\cdot\text{fm}^2$ \cite{Lyu2025}.
Then the energy shift of the $J/\psi$ immersed in a nuclear matter can be estimated to the leading order in density as
\begin{eqnarray}
   \Delta E_V= \int_{r_{min}}^{r_{max}} d^3x V(r) \rho(r) 
\end{eqnarray}
where $\rho(r)$ is the nucleon density.

We take $r_{min}$=1.0~fm because two-pion exchange no longer dominates at shorter separations. A uniform normal nuclear density of $\rho_0$=0.17~$\text{fm}^{-3}$ is used. The HAL-QCD lattice analysis shows that the $N-J/\psi$ potential vanishes within statistical uncertainties for $r>1.8$ fm \cite{Lyu2025}, so we set $r_{max}$=1.8~fm. Under these conditions, the spin-averaged energy shift is given below. 
\begin{align}
    \Delta m_{J/\psi}^{TPE}=-5\,\text{MeV}. \label{tpe result}
\end{align}

\subsection{QSR Results and Discussion}\label{sec:qsr_result}
\begin{figure}[ht]                
  \centering                      \includegraphics[width=\linewidth]{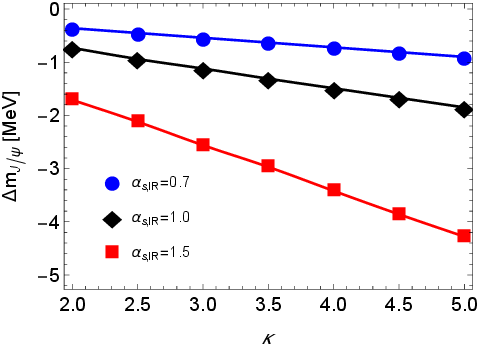}
  \caption{The mass shifts of  $J/\psi$ as a function of $\kappa$. We present results for larger infrared strong coupling constants, $\alpha_{s,\text{IR}}$.}
  \label{fig:mass shift1}         
\end{figure}

We evaluate the mass shifts of $J/\psi$ as a function of $\kappa$ as shown in Figure \ref{fig:mass shift1}. When $\alpha_{s,\text{IR}}=0.7$, although the values are smaller than that obtained from the TPE potential Eq.(\ref{tpe result}), they still display an attractive effect, with a mass decrease of 0.36--0.90 MeV. However, if the mass shift is calculated without performing the Fierz transformation, one finds a repulsive interaction. Thus, even qualitatively, extracting the pion-coupled component reproduces the attractive behavior seen in the lattice results \cite{Lyu2022,Lyu2025}.
In the analysis, we consider $\kappa$ values in the interval 2–5 to cover a physically relevant range in nuclear matter.
This is motivated by the following considerations.
The four-quark operator appearing in Eq.(\ref{scalar four quark}) contains a chirally symmetric component that is effectively absent in the vacuum saturation limit.
The magnitude of this contribution has been found to be comparable to that of the chiral breaking part \cite{Kim:2020zae,Kim:2021xyp}, implying that the four-quark matrix element in the vacuum is roughly twice as large as the vacuum-saturated value, corresponding to $\kappa = 2$.
Perturbative corrections to the Wilson coefficient of the four-quark operator, analogous to the $K$ factor in perturbative QCD, typically enhance the contribution by another factor of about two.
Taking both effects into account leads to $\kappa = 4$.
Furthermore, additional intermediate hadronic states in nuclear matter are expected to weaken vacuum dominance, and we therefore extend the range to $\kappa = 5$.
A lattice estimate of the two-pion exchange potential indicates a $J/\psi$ mass shift of order a few MeV, which serves as a reference scale; the adopted $\kappa$ values of 2–5 yield a comparable effect in this study. This parameter is not precisely known as noted previously \cite{Leinweber1997,Gubler2015,Gubler2016,Kim:2020zae,Kim:2021xyp,narison2021,JisuKim2022phi}. In Fig.\ref{fig:mass shift1}, it can be seen that the dependence on $\kappa$ is almost linear. A precise determination of $\kappa$ is essential for a quantitatively reliable estimate of the mass shift.

For the four-quark operator, we use the vacuum part coupling $\alpha_s$=0.472 at $\mu=1~\text{GeV}$ \cite{Tanabashi2018,Kim2020}, while for the density-dependent part we adopt the infrared value $\alpha_{s,\text{IR}}$=0.7 \cite{nikolaev1983,S.S.Kim2001,Deur2016}. However, the infrared value of the strong coupling constant can become larger to accommodate the long distance two-pion exchange contribution. Therefore, we evaluate the mass shifts for higher $\alpha_{s,\text{IR}}$ values, as illustrated in Fig. \ref{fig:mass shift1}. We observe that as the coupling value at long distances becomes larger, the attraction generated by two-pion exchange also increases.

As long distance physics is dominated by pions, higher order corrections to the Wilson coefficient of the four-quark operator should be important.  These effects can also be effectively parameterized into $\kappa$. Comparing our result with the lattice data suggests a larger value of $\kappa$, indicating that next-to-leading-order (NLO) corrections are important.

We also evaluate the mass shift for smaller IR couplings. We obtain that a smaller coupling yields a weaker attraction. Thus, we find that, at shorter distances that have a smaller coupling constant, the two-pion exchange contribution becomes negligible, as expected.

To examine how the long distance $N-J/\psi$ interaction depends on the heavy quark mass, we repeat the analysis with artificially larger charm quark masses. The resulting mass shift decreases as $m_c$ grows, consistent with the expectation that heavier quarkonia are more compact and therefore possess a smaller chromoelectric polarizability, which weakens long-range hadronic interactions.
The potential strength in Eq.(\ref{tpe potential}) is $\alpha_{J/\psi}=22\text{--}23~\text{MeV}\cdot\text{fm}^{2}$ \cite{Lyu2025}, while for the $N-\phi$ interaction it is $\alpha_{\phi}=91~\text{MeV}\cdot\text{fm}^{2}$ \cite{Lyu2022}. Following this trend, the potential strength parameter for the $N-\Upsilon$ interaction, $\alpha_{\Upsilon}$, is expected to be even smaller.

\section{Summary}
\label{summary}
In this work, we present the first QCD sum rules (QSR) analysis to isolate and evaluate the long-range two-pion exchange contribution to the $J/\psi$ mass shift in nuclear matter.

We obtain an attractive mass shift under the Fierz-projected treatment, which extracts the four-quark operators that couple to the two-pion exchange contribution. The sum rule results find that, although smaller in magnitude, is qualitatively in agreement with the long-range two-pion exchange potential extracted from lattice studies \cite{Lyu2022,Lyu2025}. In contrast, omitting the Fierz transformation reverses the sign of the shift, producing a repulsive interaction. Thus, isolating the pion-coupled component of the four-quark condensate is essential for capturing the correct attractive behavior seen in the lattice calculations and emphasizes the importance of properly accounting for long distance hadronic effects in QSR applications.

Furthermore, once we express the two-pion exchange contribution in terms of four-quark operators, we can identify that there are contributions from chiral symmetry breaking part, which are known to follow the change as order parameters of chiral symmetry restoration in nuclear medium. This result shows that although small in magnitude, there is a mass shift even for heavy quark system occurring due to chiral symmetry restoration in nuclear medium \cite{Lee:2000csl}.

Since two-pion exchange dominates the long distance regime, higher order corrections to the four-quark Wilson coefficient, absorbed into the parameter $\kappa$, become significant. Despite the uncertainty in $\kappa$, our analysis confirms that the $N-J/\psi$ interaction is attractive at long distances, which agrees with the lattice results.

\section*{Acknowledgements}
This work was supported by the Korea National Research Foundation under grant No. RS-2023-NR077232 and No. RS-2023-NR121103

\section{Data availability}
The data that support the findings of this article are openly available \cite{Yeo2025_Zenodo17461642}.

\appendix
\section*{Appendix}
\renewcommand{\thesubsection}{\Alph{subsection}}
\subsection{Wilson coefficient $\tilde{y}^V(\nu)$}
\renewcommand{\theequation}{A.\arabic{equation}}
\setcounter{equation}{0}
\begin{widetext}
Here, we show the explicit form of the function presented in Eq.(\ref{ope}).
    \begin{align}
        \tilde{y}^V(\nu)=\frac{\pi^2}{(4m^2)^3}\frac{\nu}{G(1/2,5/2,\nu)}&\left[-\frac{4}{3}G(1/2,7/2,\nu)-\frac
    {8}{9}G(1/2,5/2,\nu)+\frac{158}{27}G(-1/2,7/2,\nu)\right.\nonumber\\&\left.-\frac{274}{81}G(-3/2,7/2,\nu)+\frac{136}{405}G(-5/2,7/2,\nu)+\frac{2}{135}G(-7/2,7/2,\nu)\right].\label{ytilde}
    \end{align}
$G(a,b,\nu)$ is a Whittaker function which is defined as below.
\begin{align}
G(a,b,\nu)&=\frac{1}{\Gamma(b)}\int_0^\infty dte^{-t}t^{b-1}(\nu+t)^{-a}.
\end{align}

\subsection{Numerical estimation of the spin-2 four-quark condensate}
\renewcommand{\theequation}{B.\arabic{equation}}
\setcounter{equation}{0}
We estimate the numerical value of the spin-2 four-quark condensate. It is transformed as
    \begin{align}
        \langle\bar{q}_{i}\gamma_{\mu}\lambda^{a}q_{i}\bar{q}_{j}\gamma_{\nu}\lambda^{a}q_{j}|_{ST}\rangle&= \langle \bar{q}_i\gamma_{\mu}\lambda^{a}q_i\bar{q}_j\gamma_{\nu}\lambda^{a}q_j\rangle-\frac{1}{4}g_{\mu\nu}\langle \bar{q}_i\gamma_{\alpha}\lambda^{a}q_i\bar{q}_j\gamma_{\alpha}\lambda^aq_j\rangle\\
    &=-\frac{2}{3}\langle \bar{q}_i\gamma_{\mu}q_i\bar{q}_j\gamma_{\nu}q_j\rangle\nonumber-\frac{1}{8}\text{Tr}[\gamma_{\mu}\Gamma_{r}\gamma_{\nu}\Gamma_{s}]\langle \bar{q}_i\Gamma_rq_j\bar{q}_j\Gamma_sq_i\rangle \nonumber\\
    &\;\;\;\;-\frac{1}{4}g_{\mu\nu}\left(-\frac{2}{3}\langle \bar{q}_i\gamma_{\alpha}q_i\bar{q}_j\gamma_{\alpha}q_j\rangle-\frac{1}{8}\text{Tr}[\gamma_{\alpha}\Gamma_{r}\gamma_{\alpha}\Gamma_{s}]\langle \bar{q}_i\Gamma_rq_j\bar{q}_j\Gamma_sq_i\rangle \right).
    \label{nonscalarfierz}
    \end{align}

    Retaining only $\Gamma_{\mu}=\gamma_5,\gamma_5\gamma_{\mu}$, Eq.(\ref{nonscalarfierz}) becomes the following
    \begin{align}
        \rightarrow\frac{1}{4}g_{\mu\nu}\langle\bar{q}_i\gamma_5\gamma_{\mu}q_j\bar{q}_j\gamma_5\gamma_{\mu}q_i\rangle-\langle\bar{q}_i\gamma_5\gamma_{\mu}q_j\bar{q}_j\gamma_5\gamma_{\nu}q_i\rangle.\label{nonscalar}
    \end{align}
    Applying ground state dominance to Eq.(\ref{nonscalar}) in nuclear matter gives a contribution $\tfrac{9}{8}\kappa\rho_{0}^{2}$ at saturation density $\rho_0=0.17\,\mathrm{fm}^{-3}$.
\end{widetext}

\bibliographystyle{apsrev4-1}
\bibliography{refs.bib}

\end{document}